\documentclass[%
twocolumn,
superscriptaddress,
 amsmath,
 amssymb,
 aps,
 pra,
 floatfix,
 final,
 10pt,
]{revtex4-2}

\usepackage{graphicx}
\usepackage{bm}
\usepackage{datetime}
\usepackage{xcolor}
\usepackage{braket}
\usepackage{hyperref}
\usepackage[separate-uncertainty=true]{siunitx}

\renewcommand{\selectlanguage}[1]{}

\newcommand{\Erion}[1]{\mbox{\ensuremath{^{#1}}Er\ensuremath{\mathrm{^{3+}}}}}
\newcommand{\YSO}{Y$_2$SiO$_5$}
\newcommand{\Er}[1]{\ensuremath{^{#1}}Er}

\renewcommand{\Er}[1]{\ensuremath{^{#1}}Er}
\renewcommand{\Erion}[1]{\mbox{\ensuremath{^{#1}}Er\ensuremath{\mathrm{^{3+}}}}}

\graphicspath{ {figures/} } 

\usepackage[inline,]{showlabels}
\showlabels{includegraphics}
\showlabels{cite}
\showlabels{onlinecite}

\begin{document}

\author{Gavin G. G. King}\affiliation{Dodd-Walls Centre for Photonic and Quantum Technologies, University of Otago, Dunedin, New Zealand}\affiliation{Department of Physics, University of Otago, Dunedin, New Zealand}
\author{Luke S. Trainor}\affiliation{Dodd-Walls Centre for Photonic and Quantum Technologies, University of Otago, Dunedin, New Zealand}\affiliation{Department of Physics, University of Otago, Dunedin, New Zealand}
\author{Jevon J. Longdell}\email{jevon.longdell@otago.ac.nz} \affiliation{Dodd-Walls Centre for Photonic and Quantum Technologies, University of Otago, Dunedin, New Zealand}\affiliation{Department of Physics, University of Otago, Dunedin, New Zealand}

\title {Triply Resonant Microwave to Optical Conversion in Erbium-170 Doped Yttrium Orthosilicate}

\date{\today{}; \currenttime}

\begin{abstract}
We report microwave to optical upconversion in isotopically purified erbium-doped yttrium orthosilicate in a Fabry-P\'erot resonator at millikelvin temperatures. 
This follows on from investigations made at higher temperatures and with natural isotopic ratios for the erbium dopants. 
In these previous investigations the highest efficiency was seen only for moderately strong microwave powers. 
The removal of the unwanted erbium-167 which has hyperfine structure and provides unwanted background optical absorption, and the lower temperatures has removed this problem.
    We now see efficiencies still increasing as the microwave power is decreased when we reach the smallest input powers for which we could measure an output.
Efficiencies of $2\times10^{-6}$ were observed and we discuss potential improvements, including better optical cavity frequency stability and better thermalisation of the erbium spins.
  \end{abstract}


\maketitle
\section{Introduction}
\label{sec:intro}

Conversion of quantum signals from microwave to optical frequencies promises to bring many benefits to superconducting qubit quantum computing. Superconducting qubits couple to microwave photons, and so require low temperatures to operate: unless the characteristic thermal energy of the system $k_BT$ is much less than the energy of the microwave photons $\hbar\omega$, the coupled photons are swamped by thermal noise. This restriction also applies to connecting superconducting qubits; if the characteristic thermal energy of the transmission line is higher than the energy of the photons, they are quickly lost in the thermal noise. For microwave superconducting qubits, the temperatures are much less than 1\,K, and making transmission lines of any length below 1\,K quickly becomes impractical\cite{storz_loophole-free_2023}. In contrast, quantum signals encoded in optical photons can easily be transmitted over long distances, with low loss, at room temperature, using either standard optical fibers or through free space\cite{Takesue.2015, Yin.2017}.

Microwave-optical conversion would also have benefits beyond long distance communication, providing a concrete route to scaling up computer size\cite{Lecocq.2021,Awschalom.2021} and to using quantum memories for optical photons\cite{Zhong.2015,Zhou.2023}, which  are much more developed than they are for microwaves\cite{Guo.2023}.

The highest conversion efficiencies to date have been seen in Rydberg gases, with a photon-number conversion efficiency of $\eta=0.82$ \cite{Tu.2022}. Other approaches include opto-mechanical and piezo-mechanical resonators\cite{Higginbotham.2018,Jiang.2020,Zhong.2020}, electro-optics\cite{Fan.2018,sahu_quantum-enabled_2022}, and collective magnetic resonances\cite{Zhu.2020,Hisatomi.2016,Everts.2019}.

Dilute rare earth ion crystals are another exciting approach\cite{Williamson.2014,Fernandez-Gonzalvo.2015,Fernandez-Gonzalvo.2019,Xie.2021,bartholomew_-chip_2020,OBrien.2014,Welinski.2019,Bingham.1997,Bingham.1998}.  
The rare earths have narrow inhomogeneous spin and optical transition linewidths ($\sim10$\,MHz and $\sim100$\,MHz, respectively), especially at low temperatures and even with relatively high ($\sim100$\,ppm) concentrations.  
Erbium in particular has optical transitions around the 1550\,nm lowest loss window in silica fiber, and so is a natural candidate.

Raman heterodyne spectroscopy\cite{Mlynek.1983,Wong.1983} uses an optical pump to convert coherence on a spin transition to coherence on an optical transition, the decay of which produces an optical signal output that is consequently coherent with the spin transition. Using a local oscillator that is locked to the optical pump, heterodyne detection is possible, which greatly increases the sensitivity of the measurement. 

\begin{figure}
    \includegraphics[width=.5\linewidth]{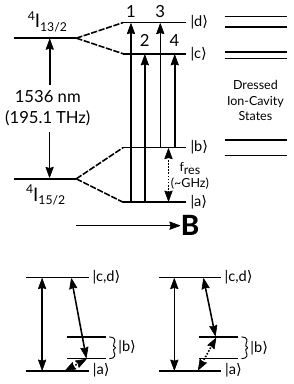}
    
    \caption{
        \label{fig:energylevels}
        The energy levels in the erbium ion used, including a sketch of the dressed states between the ions and the microwave or optical cavities.
    }
\end{figure}

Previously, conversion of  microwave photons at 4.7\,GHz to optical photons at 1536\,nm has been seen using a combined microwave and optical resonator containing a crystal of erbium-doped yttrium orthosilicate (\Erion{}:\YSO{} or Er:YSO)\cite{Fernandez-Gonzalvo.2015,Fernandez-Gonzalvo.2019}.  The conversion used three interconnected transitions shown in Figure~\ref{fig:energylevels} with a requirement for triple resonance, and was limited to $\eta=10^{-5}$ by two factors: optical absorption from unwanted \Er{167} ions, and the relatively high temperatures of the measurements. \Er{167} has a natural abundance of 22.95\%, and has hyperfine structure, unlike the even isotopes which make up the remainder of the natural composition of erbium. This hyperfine structure spans about 5\,GHz for both the $Z_1$ and $Y_1$ states and results in a complicated, broadband, poorly resolved optical and microwave absorption profile on the $Z_1\leftrightarrow Y_1$ transition\cite{Rakonjac.2020}. The comparatively high temperatures also meant that population difference between the two spin states was low and the erbium ions in the upper spin level acted to counteract those in the lower.

Later, a similar system used an isotopically purified \Er{170}:\YSO{} crystal inside a dilution refrigerator to address these problems\cite{King.2021}. There a microwave resonator was used, with a single optical pass through the sample. It showed re-absorption from the \Er{167} ion was reduced, and the microwave transition was coupled strongly to the resonator, because of the lower temperature. Raman heterodyne spectroscopy showed frequency conversion at the hybrid cavity-spin modes, the polaritons, rather than at the bare erbium frequencies.

In this paper we explore frequency conversion processes in a similar isotopically purified \Er{170} crystal, in the same combined microwave and optical resonator as used in Reference~\onlinecite{Fernandez-Gonzalvo.2019}, at temperatures below 1\,K. We find that the strong coupling seen in previously in the microwave transition is also present in the optical transition, and that Raman heterodyne signals are also seen at the hybrid optical transition-resonator frequencies. In order to minimise heating of the sample, the local oscillator was separated from the optical pump, which also allowed the local oscillator frequency to be changed, so that the measured beat signal was not at the pump frequency. This different frequency greatly reduced pick-up, increasing the sensitivity of the measurements.

\begin{figure}
    \includegraphics[width=\linewidth]{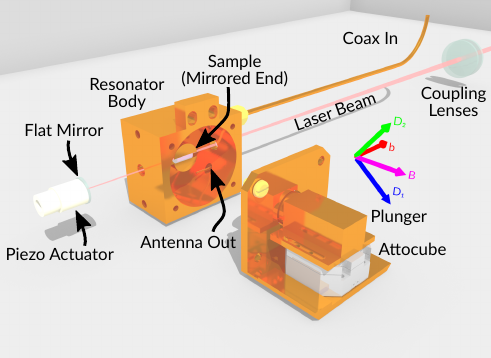}
    
    \caption{(Colour online) 
        \label{fig:layout}
        The mechanical configuration of the two resonators inside the dilution fridge, which is not shown. The microwave plunger is shown removed with its attocube. The two-axis mount for the flat mirror and piezo is not shown, but the flat mirror is in its assembled position.
    }
\end{figure}

\begin{figure*}
    \includegraphics[width=\linewidth]{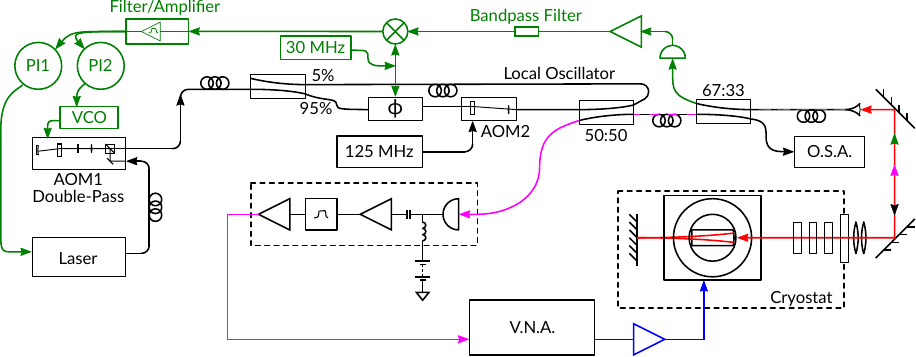}
    
    \caption{(Colour online) 
        \label{fig:signalpath}
        The signal paths for the offset-frequency conversion measurements, as described in the text. The locking system between the laser and the cavity is shown in green, the microwave input signal in blue, optical signals in red and black, and the frequency converted photons in violet. 
        O.S.A., optical spectrum analyzer; PI, proportional-integral controller; VCO, voltage controlled oscillator; V.N.A., vector network analyzer; $\phi$, phase modulator. 
    }
\end{figure*}

\section{Configuration and Methods}
\label{sec:config}

The crystal of \Erion{}-doped \YSO{} was cut from a custom-grown boule (Scientific Materials, Bozeman, Montana, U.S.A.), doped at 50\,ppm with \Erion{170}. The erbium in the precursor material was enriched to 97\% erbium-170. This isotope was chosen because the precursor had the lowest concentration of the \Er{167} ion, undesired because it has nuclear spin and hence hyperfine structure.

The sample was a cylinder 8\,mm long along its $b$ axis and  4\,mm in diameter.
The experimental apparatus comprised the crystal of \Er{}:\YSO{} inside two tunable resonators, one optical and one microwave, shown partially exploded in Figure~\ref{fig:layout}. The microwave resonator was a loop-gap resonator used in previous measurements\cite{Fernandez-Gonzalvo.2019}, where the gaps could be adjusted with an Attocube micro-positioner to tune the resonance from 5.63\,GHz to 5.90\,GHz, with $Q\approx3000$. The cavity linewidth of $\sim 2$\,MHz is well matched to typical superconducting qubit linewidths of a few megahertz\cite{kjaergaard_superconducting_2020}.  The RF magnetic field was aligned along the crystal $b$ axis. The sample was mounted in a sapphire tube to fit in the 5\,mm central hole of the resonator, which also had the effect of reducing the number of parasitic ions that were excited by the microwave field but not addressable by the optical field. 

To form the optical resonator, one end of the sample was polished to a convex radius of 107.5\,mm and coated with a dielectric high-reflectivity coating for 1536\,nm, while the other end was polished flat, and coated with an anti-reflective coating. The second end of the Fabry-P\'erot cavity was a flat high-reflectivity mirror, cut down from a commercially available mirror to reduce the mass. This flat mirror was mounted on the end of a piezoelectric stack to allow tuning. The flat mirror and piezo were attached together by a polytetrafluroethylene (PTFE) spring mount, allowing the length to change without thermal stresses breaking the piezo. The spring mount was itself mounted on a two-axis aluminium mirror mount, which was adjusted at room temperature to form the optical cavity.  The free-spectral range (FSR) of the cavity was about 2.64\,GHz, with a finesse of 310. 

The frequencies of the cavities were tuned to be resonant with each transition. The microwave cavity  has to be resonant with the microwave transition, and the optical cavity with both of the optical transitions, so that the microwave cavity frequency has to be an integer multiple of the optical cavity's free-spectral range. We used optical modes separated by two free-spectral ranges. 

Most of the beam path was within polarization-maintaining optical fiber, other than the two free-space AOMs, the fast photodetector, and coupling into the optical cavity from outside the fridge.
The output from the optical cavity was coupled into the same input fiber to return to the detector, simplifying coupling and mixing of the signal with the local oscillator.

The vibrations of the fridge and the compliance of the optical resonator meant that the optical cavity resonant frequencies varied over a range of 30\,MHz with a rate of approximately 5\,kHz.
The input fiber laser (Koheras Adjustik A820151012) was locked to an optical cavity mode by the Pound-Drever-Hall method, implemented in an FPGA device (Liquid Instruments Moku:Lab), which functioned as two separate PI controllers, as shown in Figure~\ref{fig:signalpath}.  
The laser was locked to the moving cavity because the rapid movements of the cavity piezo required to lock the cavity to the laser caused excessive heating of the sample. It was necessary to use both a slow piezo inside the laser and a fast double-pass acousto-optic modulator (AOM1).
While preferable to lock the cavity to the laser, as the frequency of the transition in the dressed state of the ion-cavity system is fixed, it was not possible because the cavity piezo was too slow to respond, as well as heating the sample when driven. 

We addressed the ions in Site~1 of the YSO crystal, with an optical transition wavelength of approximately 1536.5\,nm. Site 1 has two magnetically inequivalent orientations of \Erion{}, related by a $C_2$ rotation around the crystal $b$ axis. The effective $g$-factors of the \Erion{} ion are degenerate when the field is along the $b$ axis or is in the $D_1$--$D_2$ plane\cite{Sun.2008}.

The entire apparatus was mounted inside a home-built two-axis superconducting magnet, which was in turn mounted on a copper rod attached to the mixing chamber plate of the LD-250 dilution fridge. The magnet was wound from NbTi wire (SuperCon SC-T48B-M-0.3mm). 
The sample was oriented so that the $D_1$ crystal axis was $29.8\pm0.1^\circ$ below the axis of the magnet, as shown in Figure~\ref{fig:layout}.

There were three types of measurements made: electron paramagnetic resonance (EPR), optical absorption, and Raman heterodyne spectroscopy.  

EPR was measured by measuring the transmission through the microwave cavity with a network analyser at each applied magnetic field.

Optical absorption of the dressed state between the optical cavity and the ions was measured via the reflection off the optical cavity as the field and cavity length were varied. The laser temperature was tuned to sweep its frequency and the frequency scale was found from wavemeter measurements. 
To generate the optical spectrum shown in Figure~\ref{fig:optical:data}, the input light was modulated with an electro-optic modulator (EOM) to give three frequencies of approximately equal intensity, with the vertical scale given in terms of the central ``carrier'' frequency, with offsets set by an external microwave source to be approximately twice the optical cavity FSR, i.e. approximately the microwave cavity frequency. 

Raman heterodyne spectroscopy was measured using an ``offset-frequency'' measurement, where the local oscillator was detuned with AOM2 by 125\,MHz with respect to the pump. This means that the beat signal measured was 125\,MHz detuned from the microwave input.
This means that the desired optical signal isn't confused with electrical pickup of the microwave input signal in the detection chain.
During the Raman heterodyne measurements, the laser was locked to the optical cavity, and the frequency tuned by changing the length of the optical cavity.

We measure the conversion efficiency with the network analyser (VNA).
This is a convenient normalised measurement that can be used to easily compare different sets of parameters. To convert to a number efficiency of photons the system was calibrated. The response of the Raman heterodyne detection was measured by generating optical sidebands with the VNA and a modulator, measuring the amplitude of the  sidebands directly with the optical spectrum analyser, and comparing with the detected signal on the VNA.
To minimise differences between the measurement and calibration configurations, the phase modulator for the locking signal was repurposed to generate a ``synthetic'' sideband, and the laser detuned from the cavity so that the light reflected off the input coupling surface. 
The measured calibration was consistent with an \emph{ab initio} calibration arrived by considering each of the individual components. 

The powers were the total power incident on the cavities and were not corrected for imperfect mode matching or cavity reflection.

\section{Results}
\label{sec:results}

\begin{figure}
    \includegraphics[width=80mm]{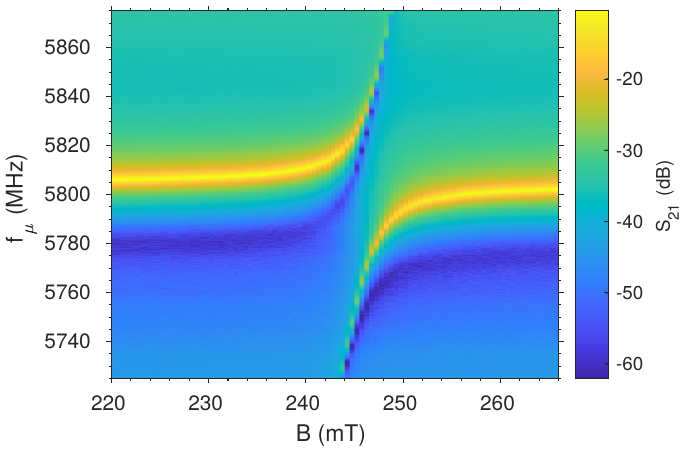}
    \caption{\label{fig:epr} (Colour online) Transmission through the microwave cavity at around 150\,mK, showing strong coupling ($\sqrt{N}g\approx 2\pi\times32$\,MHz).}
\end{figure}

Figure~\ref{fig:epr} shows the transmission through the microwave cavity at around 150\,mK as a function of magnetic field and RF frequency. There is an avoided  crossing showing strong coupling between the \Erion{170} ions and the microwave resonator. The splitting is 64\,MHz, giving a coupling strength of $\sqrt{N}g = 2\pi\times32$\,MHz. 

To double the number of effective ions, it is desirable to work with an applied field direction which maintains degeneracy between the magnetically inequivalent sub-sites, so either along the $b$ axis or in the $D_1$--$D_2$ plane.
Measuring the transmission through the microwave cavity also allowed the magnetic field to be aligned into the $D_1$--$D_2$ plane, by changing the direction of the applied magnetic field to make the resonances degenerate in frequency.

\begin{figure}
    \includegraphics[width=\linewidth]{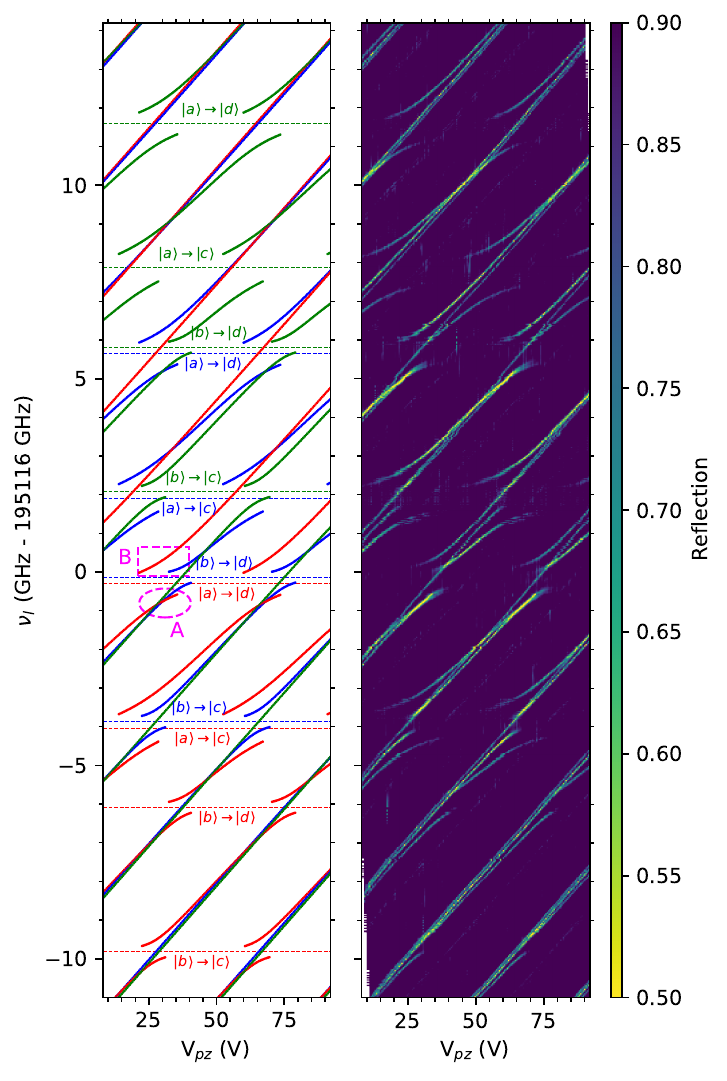}
    \caption{\label{fig:optical:data}(Colour online) The optical reflection spectrum of the resonator when three shifted frequencies are applied, using an EOM. The three frequencies are distinguished in the right-hand panel and the following figure. The transitions are as indicated by the horizontal dashed lines. The ellipse A shows overlapping dressed states, while rectangle B shows where frequency conversion was measured, as explained in the text.
    Green -- low-frequency sideband; blue -- carrier; red -- high-frequency sideband.
    }
\end{figure}

An optical reflection spectrum is shown in Figure~\ref{fig:optical:data} as a function of optical frequency and of optical cavity piezo voltage, which determines the length of the cavity. A higher voltage corresponds to a shorter cavity. The reflection off the front of the optical cavity was measured,  so that cavity resonances are seen as dips in reflection.
The figure is a composite of many individual sets of data because the laser tuning range is much less than the separation of the energy levels in the ions. 

Comparing the coupling between the like-to-like ($\ket{a}\rightarrow\ket{c}$ and $\ket{b}\rightarrow\ket{d}$) and the like-to-unlike transitions, the relative populations, and hence effective spin temperature,  can be determined.  
The effective temperature of the spins in Figures~\ref{fig:epr} and \ref{fig:optical:data} is estimated to be 350$\pm$10\,mK, somewhat higher than the copper resonator, which was measured at around 130\,mK. 

    Two sets of Raman heterodyne measurements were made: brief measurements at 4.7\,K and more detailed measurements at the base temperature of the fridge. Because the changing input powers change the frequency of the dressed cavity-ion states, changing the frequency of conversion over several linewidths, the measurements were made over a range of frequencies, and the maximum taken.

    Measurements at 4.7\,K used the optical pump as the local oscillator, allowing comparison with previous measurements of samples with natural isotopic abundance of erbium\cite{Fernandez-Gonzalvo.2019}. With an optical pump of 354\,$\mu$W and microwave input power of $-4.3$\,dBm, an efficiency of $\eta=5.6\times10^{-9}$ was seen.

\begin{figure}
\includegraphics[width=\linewidth]{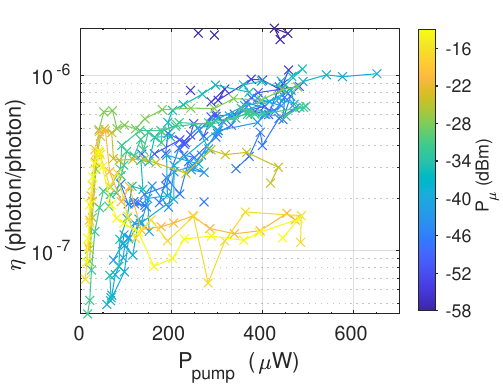}
\caption{\label{fig:rh:bothpowers} The effect of changing optical pump and microwave signal powers.}
\end{figure}

\begin{figure}
\includegraphics[width=\linewidth]{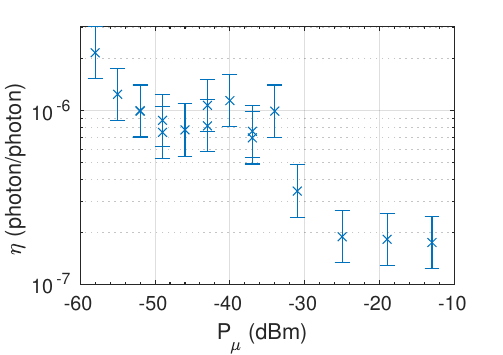}
\caption{\label{fig:rh:microwavepower} The maximum conversion efficiency seen in Figure~\ref{fig:rh:bothpowers} for an optical pump of 350--450\,$\mu$W, as the microwave signal power changes. Note that it doesn't saturate as the input microwave power falls.}
\end{figure}

    For the measurement made after cooling the  dilution fridge to base temperature, the local oscillator was separated from the optical pump and offset in frequency by 125\,MHz. The local oscillator power was fixed at $338\pm8$\,$\mu$W  for all of these measurements, while the optical pump was varied by changing the RF power driving AOM2 in Figure~\ref{fig:signalpath}. 
    We addressed the $|b\rangle\rightarrow|d\rangle$ transition in the erbium ion with the optical pump, and measured emission from the $|d\rangle\rightarrow|a\rangle$ transition, shown by the dashed rectangle {B} in Figure~\ref{fig:optical:data}. The transitions were chosen because they were the only two that had overlapping frequencies of the dressed states of the ion transitions and optical cavity modes.
    The effects of changing the optical pump and microwave input powers are shown in Figures~\ref{fig:rh:bothpowers} and \ref{fig:rh:microwavepower}. 

Figure~\ref{fig:rh:bothpowers} shows there are clearly two different types of behaviour as the optical power is increased. For high microwave powers, the conversion efficiency rapidly increased with pump power, then reached a maximum, and then fell to a constant lower level. This lower level is thought to be a result of heating from driving both transitions hard.
For low microwave powers the efficiency increased with optical power and was still increasing at the largest optical powers  used.

The lowest microwave powers have significantly fewer data points than the higher ones, because as the microwave drive decreases the signal available in the VNA for mixing down also decreases, which increased the noise floor. Therefore, it is not possible to definitely determine a trend in the lowest pump power trace, although the second lowest power shows a clear increase as the optical pump power increases.

Taking the maximum of each line between 350\,$\mu$W and 450\,$\mu$W in Figure~\ref{fig:rh:bothpowers}, a clear trend of increasing conversion efficiency can be seen in Figure~\ref{fig:rh:microwavepower}. There is a perhaps unexpected bump at around $-40$\,dBm of microwave power, which is likely a result of the sample heating due to both high optical and microwave pump powers. However, the general trend is for the conversion efficiency to increase as the microwave drive decreases, reaching a maximum number efficiency of $2\times10^{-6}$ at the lowest power measurable.

\section{Discussion}
\label{sec:discuss}

EPR shows strong coupling between the ground-state transitions in the erbium ions and the microwave cavity, which is a prerequisite for efficient frequency conversion\cite{Williamson.2014}. The coupling strength is enhanced by the low temperatures\cite{Fernandez-Gonzalvo.2019}, and is consistent with other measurements made of this material\cite{King.2021}.

  The optical spectrum in Figure~\ref{fig:optical:data} was used to help choose the pair of optical transitions to use. We require both the optical pump and generated optical signal to be resonant with an ion-cavity dressed mode, and we want those modes to have a strong mixture of ion and cavity nature.
  At these temperatures the lifetime of the transitions is very long, so the populations are very sensitive to the amount of power put into the transitions. The overlapping modes in the dashed ellipse {A} in Figure~\ref{fig:optical:data}, generated by the relatively strong sidebands of an EOM, were not the transitions used for Raman heterodyne spectroscopy. Instead, the low population of the excited $|d\rangle$ state when measuring frequency conversion meant that the two dressed states at around 0\,GHz (the dashed rectangle {B} in  Figure~\ref{fig:optical:data}) overlapped, and so were used.  
    One improvement that could be made to the measurements is a better understanding of exactly how the populations change the dressed states, so that they can be better controlled, perhaps allowing a choice of arbitrary transitions and optical modes.

  The most important measurement was microwave optical frequency conversion, which showed conversion efficiency of $\eta=5.6\times10^{-9}$ at 4.7\,K, and $\eta=2.0\times10^{-6}$ at around 350\,mK. The efficiency at 350\,mK is comparable to that measured at 4.6\,K in previous works\cite{Fernandez-Gonzalvo.2019}, but with an optical pump an order of magnitude less. The reduced temperature causes some of this improvement, but a large part can be attributed to the elimination of parasitic re-absorption, by eliminating the \Erion{167} ion.
  One significant improvement over Reference~\onlinecite{Fernandez-Gonzalvo.2019} is that good efficiencies are seen in the limit of low microwave powers. Indeed, now the efficiency increases as we lower the microwave power (Figure~\ref{fig:rh:microwavepower}).

Additionally, parasitic absorption of the input microwave photons was reduced by the sapphire tube used to mount our sample. It was not the primary purpose of the sapphire tube, which was used because the sample available was smaller in diameter than the previous sample. There were fewer erbium ions outside of the optical mode volume, so that there were fewer of the so-called parasitic ions that absorb a microwave photon, but can't possibly be pumped to emit an optical photon.

A particular limiting factor was the thermalisation of the  sample. There was a large thermal gradient between the resonator with the sample (typically 150--250\,mK) and the mixing chamber of the dilution fridge (60--120\,mK), caused by the thin cold finger between the two. The effective temperature of the spins could be inferred from the optical measurements of thermal populations, and was found to be higher still, estimated to be 350\,mK, which is high enough to cause some thermal excitation of the microwave transition, and so limit the conversion efficiency by limiting the population difference between the microwave levels.  

In order to reduce heating of the sample, pulsed operation is desirable, where the pump is modulated such that the instantaneous power is high, but the average power is not.   For the scheme in Figure~\ref{fig:signalpath}, the pump light was used to lock the laser to the cavity. With the unstable cavity it is very difficult to construct a scheme such that the cavity remains locked as the cavity jitters while the pump power changes by perhaps orders of magnitude, or is turned on and off.

Mechanical instability had a big effect on our ability to observe high conversion efficiencies. 
The jitter in the optical cavity effectively pulsed the pump laser and therefore the signal, which meant that the conversion process was only happening when the jitter brought the optical cavity into alignment. This is one potential reason why we saw a conversion efficiency much less that Reference~\onlinecite{Fernandez-Gonzalvo.2019} at 4.7\,K, although our 4.7\,K measurement was very preliminary and little effort was put into optimising the efficiency.
Broadening of the pump can be accounted for by increasing the bandwidth of the detection, at the cost of decreasing the signal to noise ratio, but the pump light lost due to direct reflection off of the cavity directly impacts the efficiency. 

A particular improvement to be made is to improve the stability of the optical cavity.  Possibilities include: making the alignment of the two mirrors more stable and rigid; using a fiber inside the fridge to couple to the resonator; or using a monolithic cavity\cite{ma_optically_2023}, which would reduce the mechanical instability, at the expense of no longer having tunable optical modes.

Modeling of a similar system\cite{barnett_theory_2020} suggests that an overall efficiency of over 50\% is conceivable with \Erion{}:\YSO{}. While these measurements are not close to that efficiency, we can make some estimates about the expected improvements from what was measured. By decreasing the temperature from 4.7\,K to 250\,mK, we saw an increase of efficiency of roughly $10^3$. If we were able to reproduce the efficiency of $\eta=10^{-5}$ \cite{Fernandez-Gonzalvo.2019} previously seen at 4\,K, then the possible efficiency at 350\,mK could reach $10^{-2}$. 
By better thermalising the system, perhaps by pulsing the pump light, and reducing the temperature to 50\,mK, we would gain another order of magnitude, so that an overall efficiency of $\eta\sim10^{-1}$ is plausible.

\section{Conclusion}
\label{sec:conclusion}

We have measured frequency conversion from microwave photons to optical photons in a rare earth ion system, showing a maximum total conversion efficiency of $\eta=2\times10^{-6}$. Unlike previous measurements, this system used an isotopically purified sample, reducing parasitic re-absorption from hyperfine structure. The system showed a similar conversion efficiency to the natural abundance sample at 4\,K, but with a much lower optical pump power. By further reducing the temperature the conversion efficiency increased, as expected by the theory. The conversion efficiency increased with decreasing microwave input power, towards the desired single photon regime for quantum computing interconnects, and with increasing optical pump power. The conversion efficiency was limited by thermalisation of the sample (the effective temperature of the spins was much higher than that of the dilution fridge they were mounted in),  by the mechanical stability of the optical cavity, and by the noise limits of the detection apparatus, so that no maximum was observed, suggesting the system was not saturating. It is therefore likely that higher efficiencies can be seen with this system, perhaps approaching the 50\% efficiency that is necessary for certain quantum error correction algorithms to function.

While preparing this manuscript, interesting results have been reported\cite{xie_scalable_2024} using  \textsuperscript{171}Yb:YVO\textsubscript{4}\cite{bartholomew_-chip_2020}, which benefits immensely from an unusually narrow spin-transition with inhomogeneous linewidth of 160\,kHz, due to magnetic insensitivity.

\begin{acknowledgments}

This work was supported by the U.S.A.\ Army Research Office (ARO/LPS) (CQTS) Grant No. W911NF1810011, and  the Marsden Fund (Contract No.  UOO1520) of the Royal Society of New Zealand.
\end{acknowledgments}

\bibliography{gkrefs,lstrefs}

\end{document}